%% file: ring.tex
\documentstyle[epsfig]{europhys}
\input euromacr

\begin{document}
\euro{?}{?}{1-$\infty$}{1999}
\Date{? ? 1999}
\shorttitle{T. Winiecki {\it et al.}, Vortex structures}
\title{Vortex structures in dilute quantum fluids}
\author{T. Winiecki\inst{1}, J. F. McCann\inst{2},
\And , C. S. Adams\inst{1}}
\institute{
     \inst{1} Dept. of Physics, University of Durham, Rochester Building,
South Road, Durham, DH1 3LE, England. UK.\\
     \inst{2} Dept. of Applied Mathematics and Theoretical Physics, 
Queen's University, Belfast, BT7 1NN, Northern Ireland. UK.}
\rec{}{}
\pacs{
\Pacs{02}{30}{Nonlinear differential equations}
\Pacs{03}{75.F}{Bose-Einstein condensation}
\Pacs{47}{37.+q}{Superfluid hydrodynamics}
\Pacs{67}{40.Vs}{Vorticity in $^4$He}}
\maketitle

\begin{abstract}
Vortex structures in dilute quantum fluids are studied using the Gross-Pitaevskii 
equation. The velocity and momentum of multiply quantized vortex rings are 
determined and their core structures analysed. For flow 
past a spherical object, we study the encircling and pinned ring solutions,
and determine their excitation energies as a function of the flow velocity
for both penetrable and impenetrable objects. The ring and laminar
flow solutions converge at a critical velocity, which decreases with 
increasing object size. We also study the vortex solutions
associated with flow past a surface bump which indicate that
surface roughness also reduces the critical velocity. 
This effect may have important implications for 
the threshold of dissipation in superfluids and superconductors.

\end{abstract}

\section{Introduction}
The breakdown of superfluidity in liquid helium (HeII) \cite{donn91,till90}, 
dissipation in superconductors \cite{till90}, and topological defects in 
cosmology \cite{hind95} all involve vortices. Given their prevalence, 
it is perhaps surprising how little is known about their structure and 
the exact mechanism of their formation. For example, in HeII 
vortex rings can be produced experimentally by injecting ions into the 
fluid \cite{donn91,till90}. However, it is not known whether the rings 
emerge via {\it quantum transition}, where the ion creates an encircling 
ring and subsequently attaches itself to the vortex core; or {\it peeling}, 
where the ion creates a vortex loop which grows to form a pinned ring 
\cite{donn91,till90}. In principle, this question could be addressed by 
simulation, except that a complete hydrodynamical model of HeII has yet 
to be established. One can formulate a simple model of a dilute Bose 
fluid based on the Gross-Pitaevskii or non-linear Schr\"odinger 
equation (NLSE) \cite{gpe}. This approach can be expected to 
provide an accurate quantitative description in the case of dilute 
Bose-Einstein condensates \cite{dalf99}, and offer considerable
insight into the physics of HeII. Vortex solutions of the 
NLSE equation have been studied for a number of simple geometries:
Flow past an object has been studied in one \cite{hak97} and two 
dimensions \cite{huep97}, and it is found that stationary vortex solutions 
exist only for motion slower a critical velocity. Above 
this velocity, one observes the periodic emission of vortices leading 
to a pressure imbalance which produces drag on the object \cite{wini99}. 
Consequently, the critical velocity also determines the transition between 
superfluid and normal flow.\\

In this paper, we investigate stationary vortex solutions near an object 
in three dimensions. For a spherical object, there are two solutions:
the {\it encircling ring}, where the object is positioned at the centre of the
ring; and the {\it pinned ring}, 
where the object centre lies within the core of the vortex line. 
For all object potentials laminar flow evolves smoothly into
the encircling ring solution suggesting that the quantum
transition model of ring nucleation is favoured. Finally,
we study flows parallel to a plane and illustrate the effect of
surface roughness on the critical velocity.

\section{Numerical method}
Consider an object described by a potential $V(\mbox{\boldmath $r$})$ 
moving with velocity $\mbox{\boldmath $v$}$ through a fluid of interacting 
bosons with mass $m$ and uniform asymptotic number density $n_0$.
In the fluid rest frame,
$\mbox{\boldmath $r$}'= \mbox{\boldmath $r$}+ \mbox{\boldmath $v$}t$,
the dynamics can be described in terms of the order parameter, 
$\psi(\mbox{\boldmath $r$}',t)$, which is a solution of the NLSE,
\begin{equation}
i\hbar {\partial \over \partial t}\psi(\mbox{\boldmath $r$}',t)
=-{\hbar^2 \over 2m}\nabla'^2 \psi(\mbox{\boldmath $r$}',t)+
V(\mbox{\boldmath $r$}'-\mbox{\boldmath $v$}t)\psi (\mbox{\boldmath $r$}',t)+
C \vert\psi(\mbox{\boldmath $r$}',t)\vert^2 \psi(\mbox{\boldmath $r$}',t)~,
\label{eq:nlse}
\end{equation}
where the nonlinear coefficient, $C=4\pi\hbar^2a /m$, and $a$ is the $s$-wave
scattering length. If distance and velocity are measured in terms of 
the healing length, $\xi=\hbar/\sqrt{mn_0C}$, and the speed of sound, 
$c=\sqrt{n_0C/m}$, respectively, and the asymptotic number density is rescaled 
to unity, then Eq.~(\ref{eq:nlse}) in the frame of the object becomes
\begin{equation}
i \partial_t \psi=-\textstyle{1\over 2}\nabla^2\psi+V\psi +\vert\psi\vert^
2 \psi+i \mbox{\boldmath $v$}\cdot\nabla \psi~.
\label{eq:nlse2}
\end{equation}
Stationary states,
$\psi(\mbox{\boldmath $r$},t)={\rm e}^{-i\mu t}\phi(\mbox{\boldmath $r$})$,
corresponding to a chemical potential, $\mu=1$, are found by numerical solution 
of the discretized system of equations. In the special case of rotational 
symmetry about the axis of flow, the problem can be solved using 2D cylindrical 
polar coordinates $(\rho,z)$. In the general case, we use a cartesian 3D grid 
$\phi(\mbox{\boldmath $r$})=\phi(x,y,z)$. Defining 
$\phi_{ijk0}=\Re(\phi(x_i,y_j,z_k))$ and $\phi_{ijk1}=\Im(\phi(x_i,y_j,z_k))$, 
and taking the flow direction to define the $z$-axis, Eq.~(\ref{eq:nlse2}) becomes
\begin{eqnarray}
\label{eq:nlin}
f_{ijkr}  \equiv  -\left(\phi_{i+1 j k r}+\phi_{i-1 j k r}
+\phi_{i j+1 k r}+\phi_{i j-1 k r}+\phi_{i j k+1 r}+\phi_{i j k-1 r}
-6 \phi_{i j k r} \right)/2 \Delta^2\\ \nonumber
+  V_{ijk} \phi_{ijk r}+  \left( \phi_{ijk0}^2+ \phi_{ijk1}^2\right) \phi_{ijkr}
-  \phi_{ijkr}+(2r-1) v (\phi_{ijk+1,1-r}-\phi_{ijk-1,1-r})/2\Delta =0
\end{eqnarray}
where $\Delta$ is the grid spacing (values of $\Delta$ between 0.125 and 0.4 were 
used). The solution of these equations is found using the linearization
\begin{equation}
f_{ijkr} ( \phi_{lmns}^{(p)}) + \sum_{lmns}
 \left( \phi_{lmns}^{(p+1)} - \phi_{lmns}^{(p)} \right)
\left[ {\partial f_{ijkr} \over  {\partial \phi_{lmns}}} \right]^{(p)} \approx 0~,
\label{eq:lin}
\end{equation}
where $\phi^{(p+1)}$ is determined from the approximation $\phi^{(p)}$ by solving Eq.~(\ref{eq:lin}) using the bi-conjugate gradient method \cite{numrec}. Large
box sizes ($0\leq x\leq \infty$) are simulated by employing the nonlinear 
transformation $\hat{x}=x/(x+D)$, where $D$ is chosen to give sufficient 
point coverage within the healing length. The iterative solution depends 
on the initial guess $\phi^{(0)}$; for example, laminar flows are found by 
choosing $\phi^{(0)}=\mu=1$, whereas vortex solutions are found
by imposing the vortex phase pattern on $\phi^{(0)}$, and then relaxing
the imposed phase after a few iterations.

\section{Free vortex rings}
For an unbounded flow ($V=0$) one finds free vortex rings. Fig.~1 shows a surface 
plot of ring solutions with circulation, $\kappa=2\pi$, $4\pi$, and $6\pi$.
The double ($\kappa=4\pi$) and triple ($\kappa=6\pi$) rings only exist for 
$v>0.56$ and $v>0.67$, and their cores consists of 2 and 3 lines of zero
density, respectively. The separation of the density minima depends on the radius, increasing up to about two healing lengths for $R=5$. For $\kappa=6\pi$, 
the central minima has a larger radius. Similar core structures are also found
in the corresponding 2D solutions. Although vortex rings with multiple circulation 
have higher energy ($E\propto\kappa^2$) than the corresponding number of single rings
($E\propto\kappa$), we find that they are stable when subject to a perturbation 
in a time-dependent simulation (similar robustness has also been
found for vortex lines with multiple circulation \cite{aran96}).
\begin{figure}[htb]
\epsfig{file=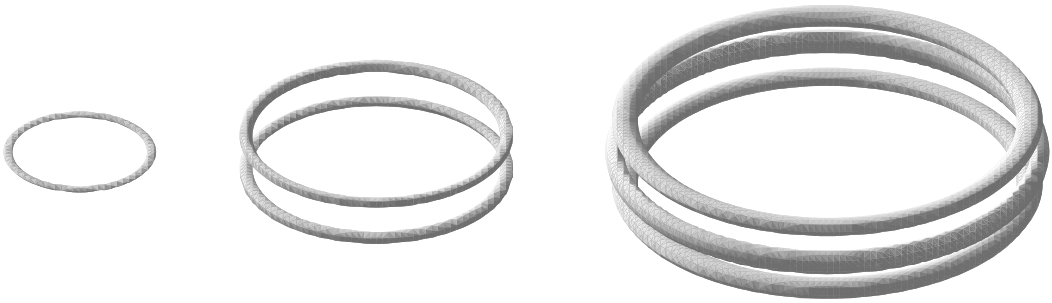,clip=,width=14.0cm,bbllx=130,bblly=370,bburx=465,bbury=470}
\caption{Surface density ($\vert \phi \vert^2=0.02$) of vortex rings with 
$\kappa=2\pi$, $4\pi$, and $6\pi$ for $v=0.5$. The flow direction is parallel to 
the axis of the rings. The core structure is only visible at very low density:
the surface density plot for $\vert \phi \vert^2=0.08$ appears as a single ring
in each case.}
\label{fig:1}
\end{figure}
Fig.~2 shows a plot of the velocity (equivalent to the ring propagation 
velocity in a stationary fluid) and momentum as a function of ring radius, $R$. 
\begin{figure}[hbt]
\centering
\epsfig{file=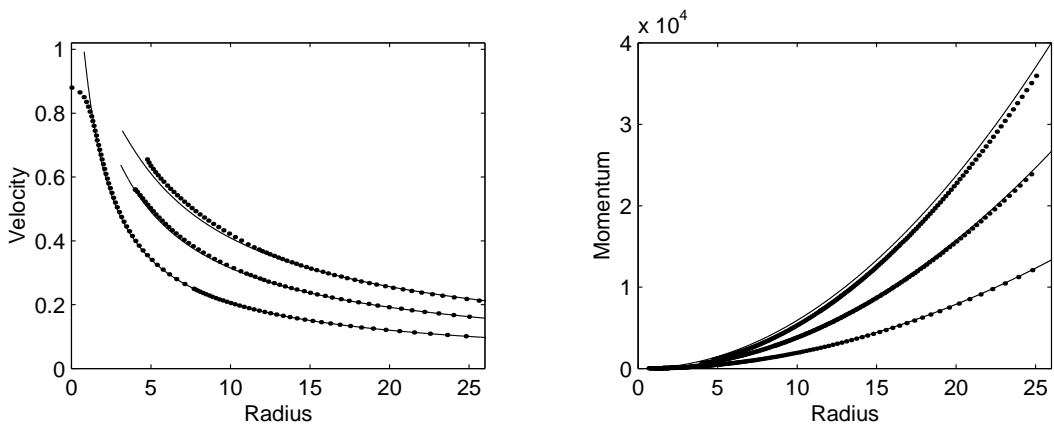,clip=,width=14.0cm,bbllx=60,bblly=200,bburx=540,bbury=395}
\caption{Vortex ring velocity (left) and momentum (right) for $\kappa=2\pi$, 
$4\pi$, and $6\pi$: The numercial data ($\cdot$) are
compared to $v=\kappa\left[\ln(8R/a)-b\right]/4\pi R$ ($a=1/\sqrt{2}$, 
$b=0.615$, 1.58, and 2.00), and $P= \pi \kappa R^2 $, respectively.}
\label{fig:2}
\end{figure}
For $R>5$, the velocity can predicted accurately by 
$v=\kappa\left[\ln(8R/a)-b\right]/4\pi R$, where $a$ is the core
radius ($a=1/\sqrt{2}$ in our units) and $b$ is a constant which depends
on the structure of the core ($b=0.25$ for a classical fluid, whereas
$b=0.615$ has been predicted for a dilute quantum fluid \cite{rob71}).
Numerical fits give $b=0.615$ (as expected), 1.58, and 2.00 for $\kappa=2\pi$,
$4\pi$ and $6\pi$, respectively. For $\kappa=2\pi$ (single ring), the ring 
disappears (i.e. the on-axis density becomes zero) when $v=0.88$, however, 
a solution other than laminar flow exists up to $v=1.0$ (see below).

\section{Flow past a sphere}
The free ring solutions discussed above are modified by the presence of an
object or surface. In this section, we present results for a spherical object 
with radius $R$ and potential height $V$, i.e., $V(r)=V$ ($r \leq R$) 
and $V(r)=0$ ($r>R$). Fig.~3 shows surface density images of the three 
possible solutions: {\it laminar flow}; the {\it pinned ring} or {\it vortex loop};
and the {\it encircling ring} (the corresponding 2D solutions:
laminar flow; a free and a bound vortex, and a vortex pair, were studied 
by Huepe and Brachet \cite{huep97}).
\begin{figure}[htb]
\centering
\epsfig{file=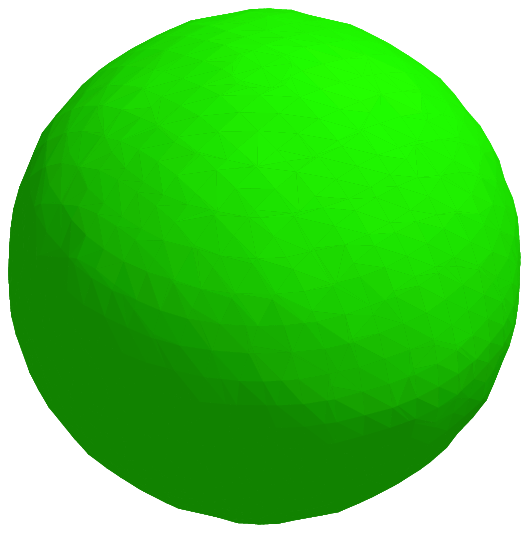,clip=,width=4.65cm,bbllx= 70,bblly=250,bburx=520,bbury=530}
\epsfig{file=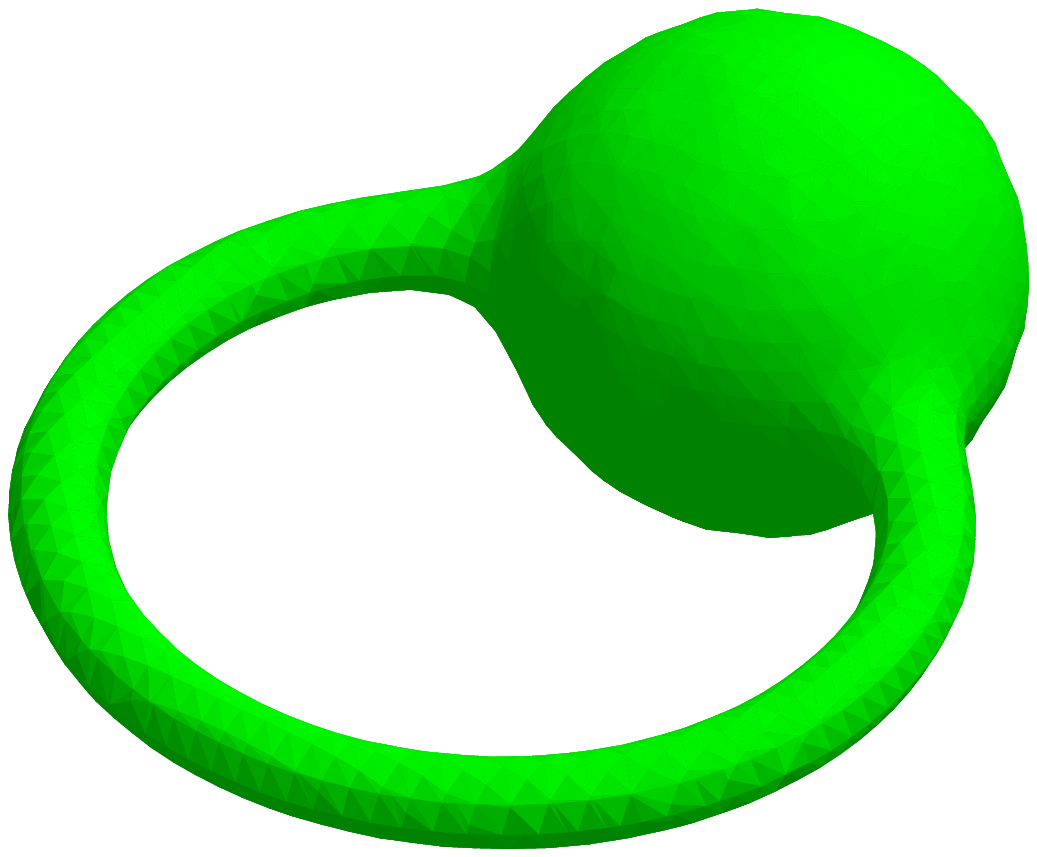,clip=,width=4.65cm,bbllx=130,bblly=300,bburx=580,bbury=580}
\epsfig{file=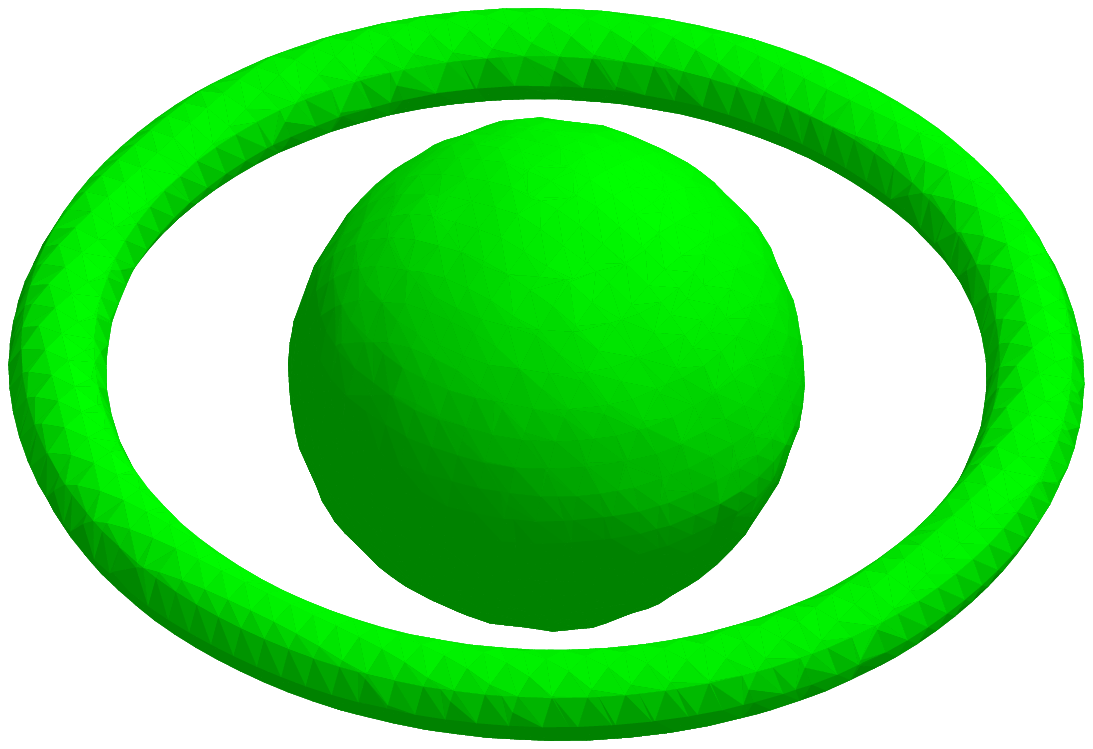,clip=,width=4.65cm,bbllx= 70,bblly=250,bburx=520,bbury=530}
\caption{Surface density plots ($\left| \phi \right|^2=0.25$) showing the three 
steady-state solution associated with flow past a spherical object: from left to
right, laminar flow; pinned ring; encircling ring.}
\label{fig:3}
\end{figure}
Fig.~\ref{fig:4} shows a section of the velocity field pattern around the obstacle for
each case. As the flow velocity increases, the vortices move closer to the object and
eventually merge into the surface. Close to the critical velocity,
the flow patterns converge (Fig.~\ref{fig:4} lower).
\begin{figure}[htb]
\epsfig{file=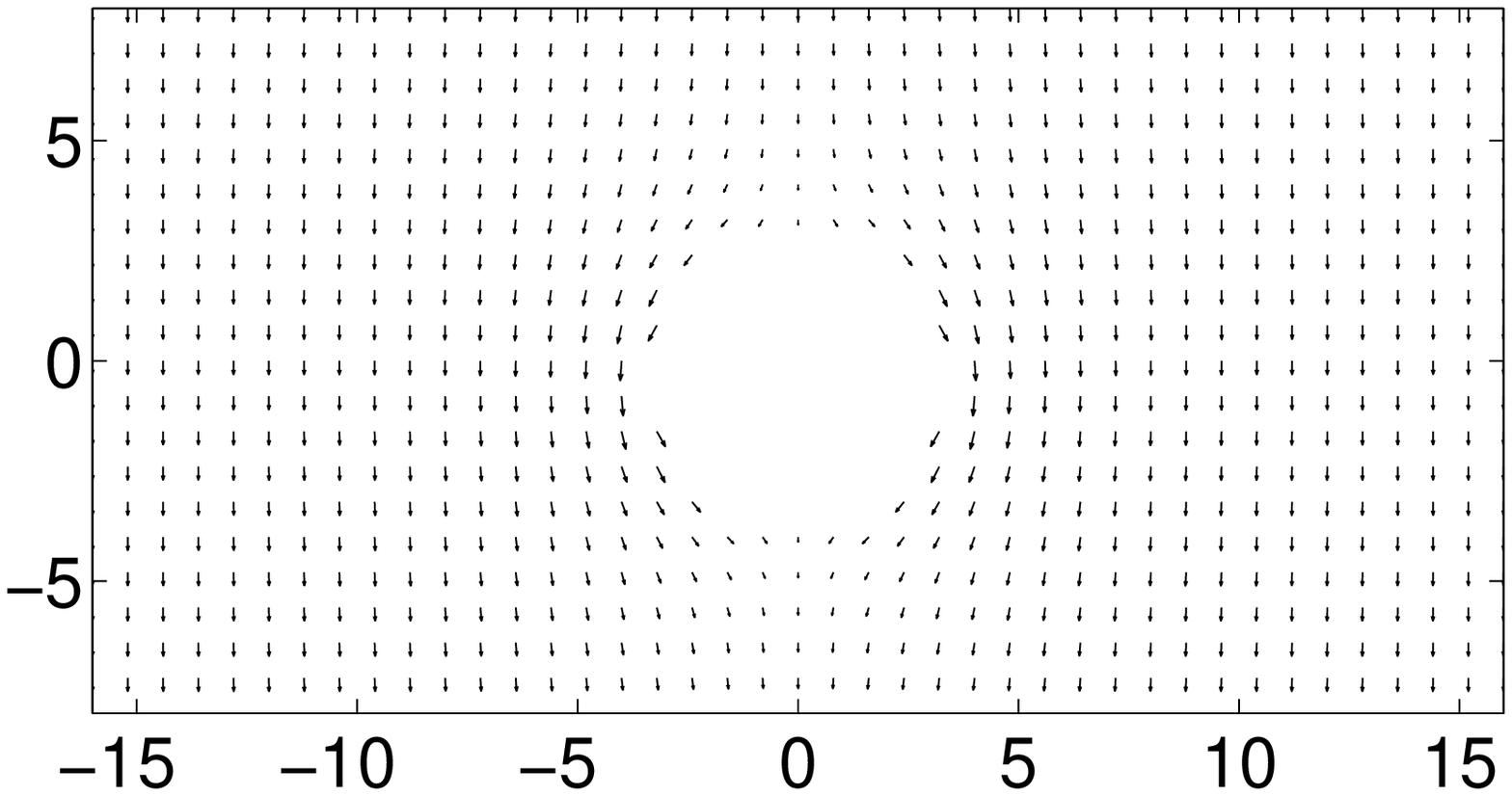,clip=,width=4.87cm,bbllx=60,bblly=290,bburx=542,bbury=520}
\epsfig{file=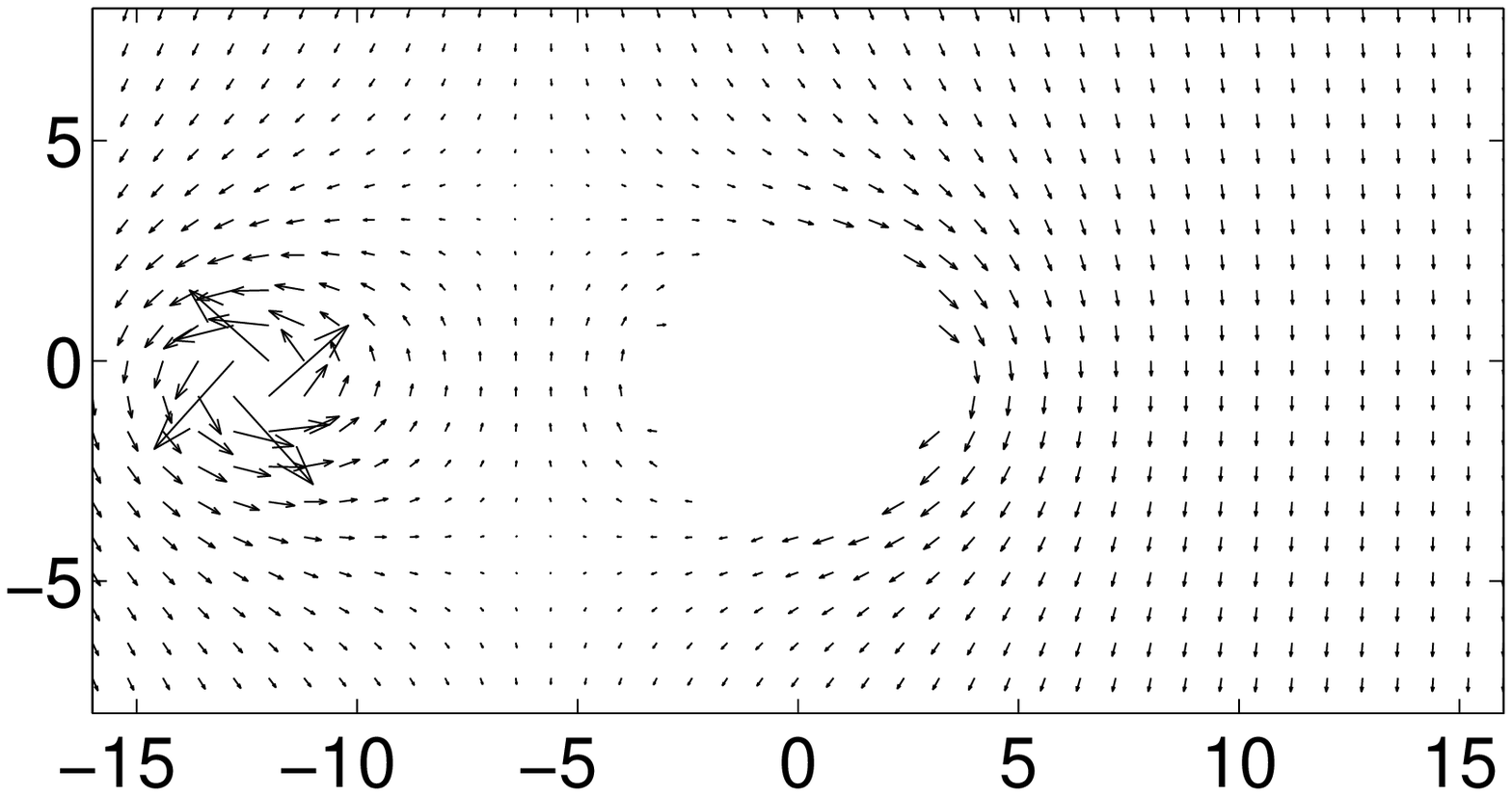,clip=,width=4.57cm,bbllx=90,bblly=290,bburx=542,bbury=520}
\epsfig{file=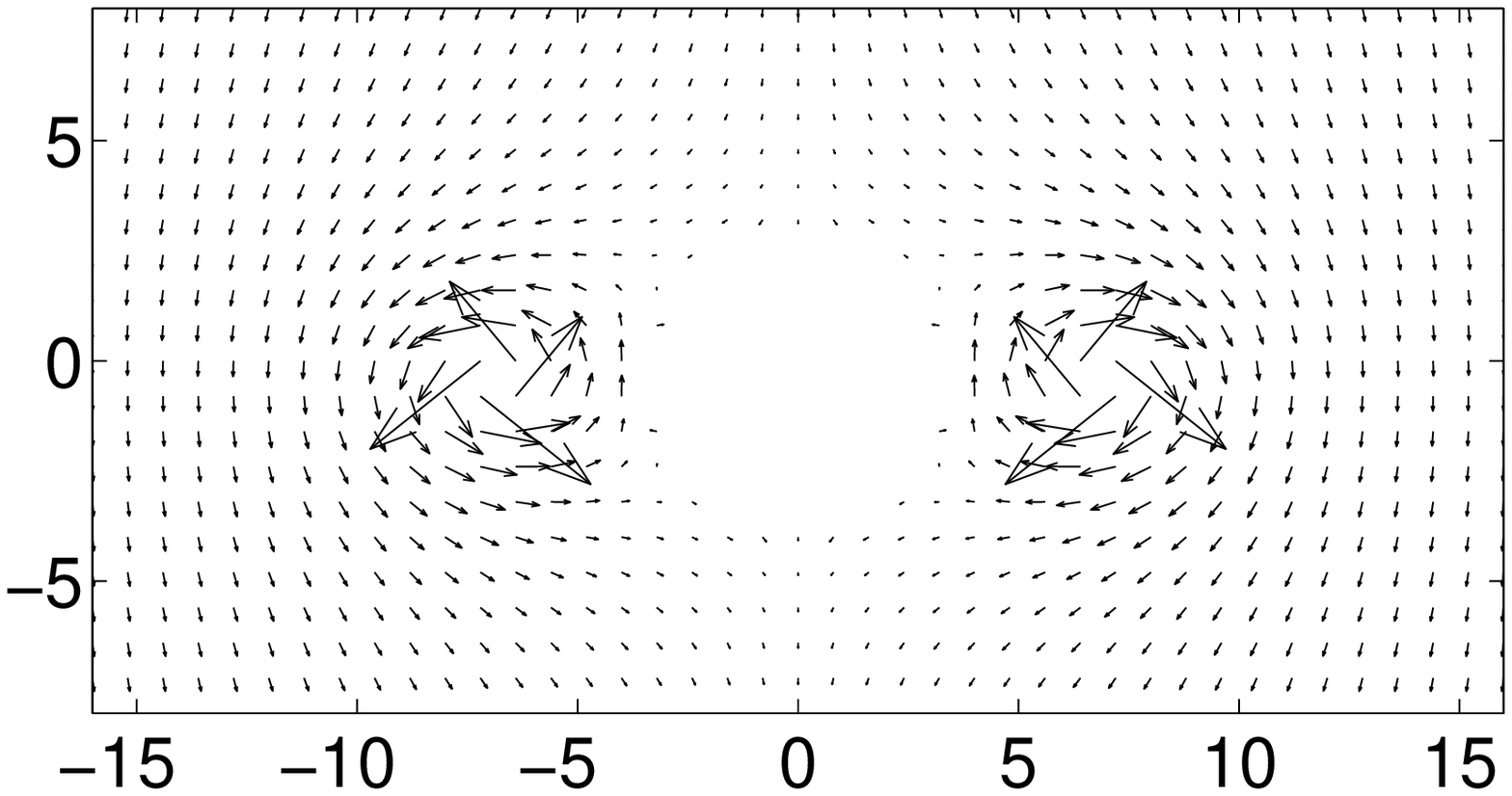,clip=,width=4.57cm,bbllx=90,bblly=290,bburx=542,bbury=520}
\\
\epsfig{file=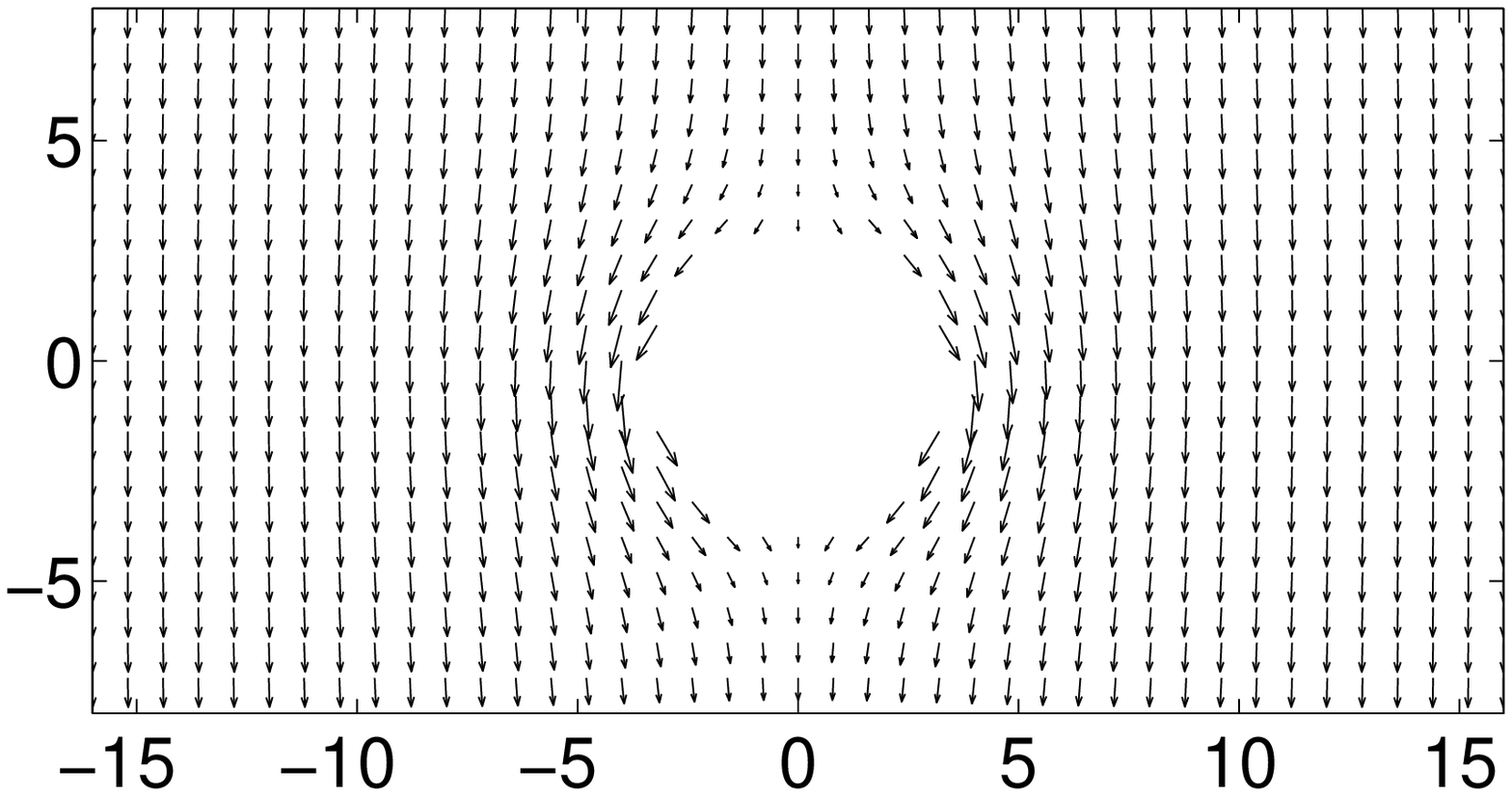,clip=,width=4.87cm,bbllx=60,bblly=265,bburx=542,bbury=520}
\epsfig{file=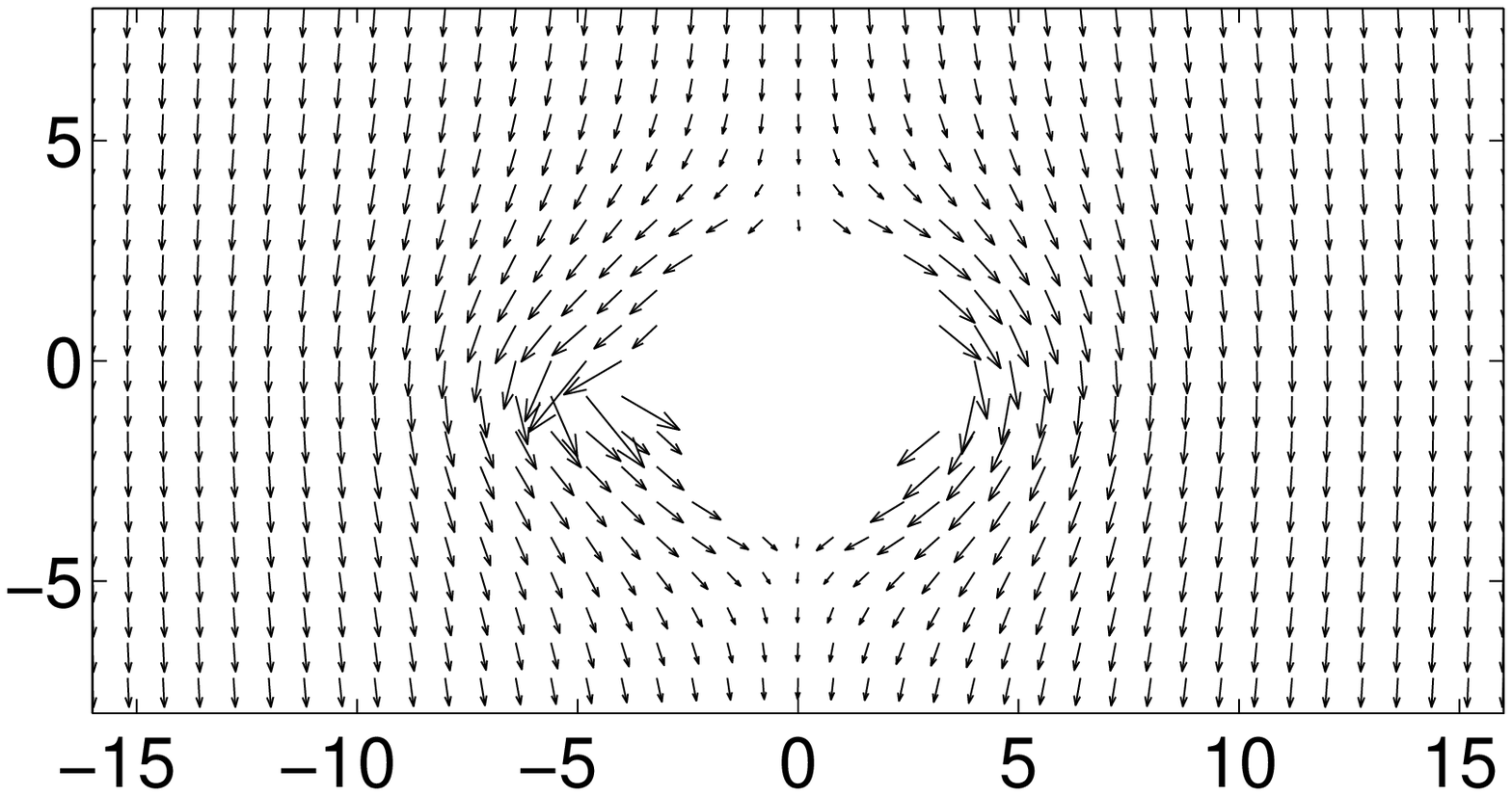,clip=,width=4.57cm,bbllx=90,bblly=265,bburx=542,bbury=520}
\epsfig{file=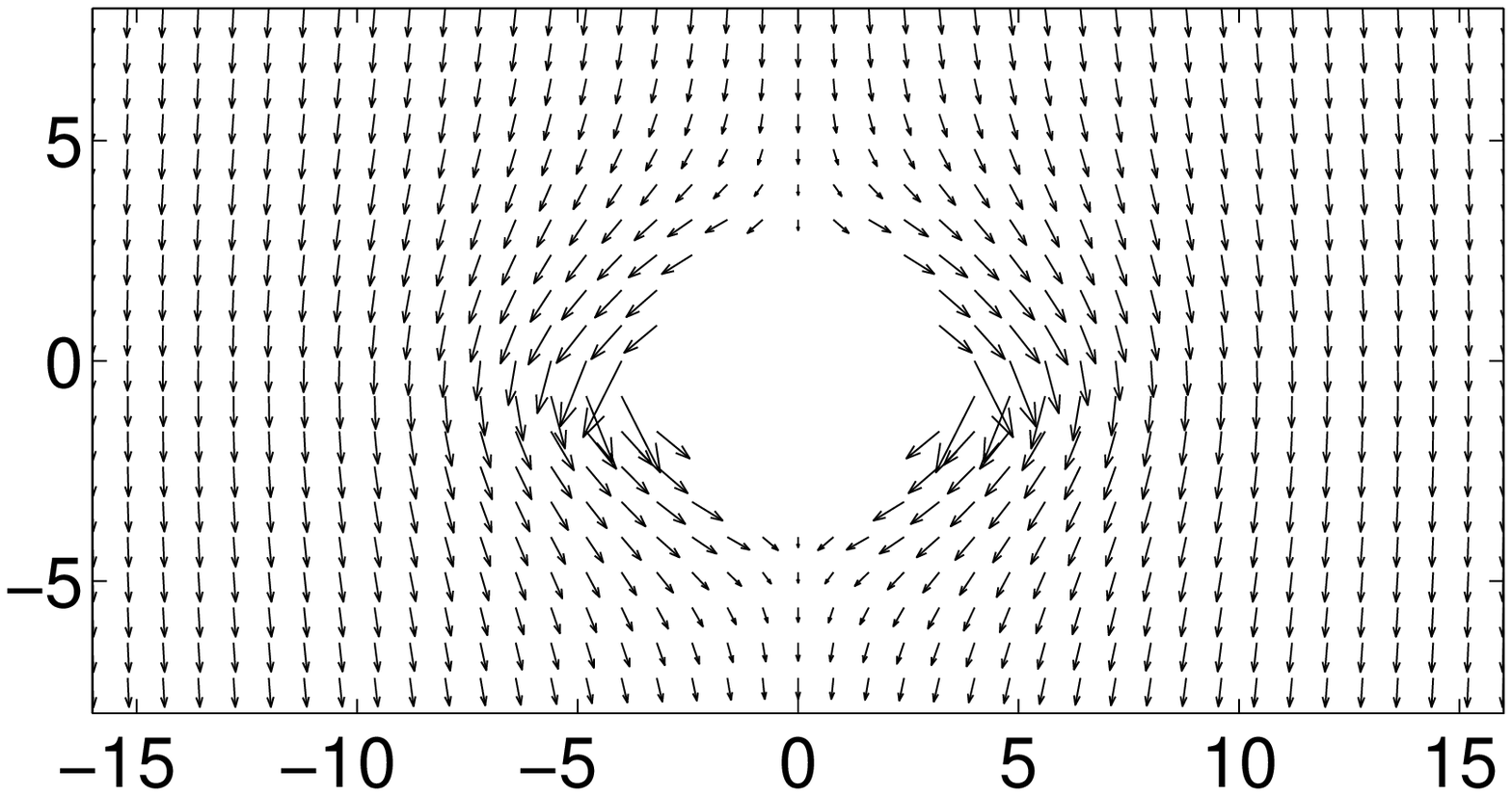,clip=,width=4.57cm,bbllx=90,bblly=265,bburx=542,bbury=520}
\caption{The velocity field pattern in the plane of a spherical object
with radius $R=3.3$ for $v=0.3$ (above) and $v=0.66$ 
(below) (the critical velocity, $v_{\rm c}=0.678$). The three columns
correspond to laminar flow (left), the pinned ring (middle), and the 
encircling ring (right). Note the circulation around the object in the 
pinned ring solution. Only one quarter of actual grid points are shown.}
\label{fig:4}
\end{figure} 

\section{Vortex energy}
\begin{figure}[htb]
\centering
\epsfig{file=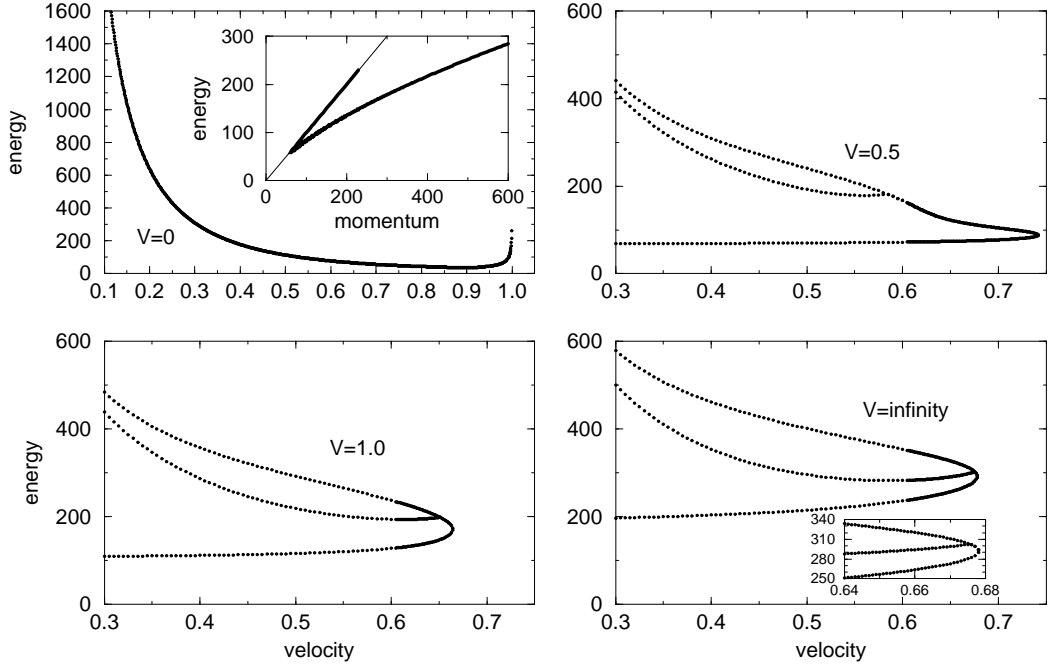,clip=,width=14.0cm,bbllx=60,bblly=45,bburx=510,bbury=330}
\caption{The energy of the stationary solutions of the NLSE as a function of
the flow velocity. With no obtacle (upper left) `ring' solutions are found
up to the speed of sound $v=1.0$. The ring collapses at $v=0.88$
leaving a localised solitary wave with energy, $E\sim cP$ (see inset). If a sphere 
($R=3.3$) is inserted in the flow, the ring solution splits into a 
pinned and an encircling ring (higher energy). 
Even for $V=\infty$, the encircling and pinned solutions merge 
below the critical velocity (inset, lower right). }
\label{fig:5}
\end{figure}

The Lagrangian density of the NLSE (\ref{eq:nlse}) in scaled units, is given by
\begin{equation}
{\cal{L}}(\psi,\nabla\psi,\dot{\psi})=i\psi^* \dot{\psi}-
{1\over2}\nabla \psi^*\nabla\psi-V\psi^*\psi
-{1 \over 2}\left( \psi^* \psi \right)^2~. 
\end{equation}
From the Hamiltonian, $H= \int {\rm d}\mbox{\boldmath $r$} 
[(\partial {\cal{L}}/\partial \dot{\psi})\dot{\psi}-\cal{L}]$,
we define the energy relative to the ground state  
(i.e. laminar flow with $V=0$ and having the same number of particles), as
\begin{equation}
E=\int {\rm d} \mbox{\boldmath $r$} \left\{ {{1 \over 2} \left| \nabla 
\phi \right|^2+ V \left| \phi \right|^2 +{1 \over 2} \left( \left| 
\phi\right|^2 -1 \right)^2 }\right\}~.
\label{eq:energy}
\end{equation}
Any deviation of the local particle density $ \left| \phi\right|^2 $ from $1$ 
constitutes an excitation. Eq.~(\ref{eq:energy}) can be rewritten as
\begin{equation}
E= \int {\rm d} \mbox{\boldmath $r$} \left\{ {1 \over 2} \left(
1 -\left| \phi \right|^4\right ) \right\} +\mbox{\boldmath $v$} 
\cdot \mbox{\boldmath $P$}~,
\label{eq:ener2}
\end{equation}
where $\mbox{\boldmath $P$}=-i\int {\rm d}\mbox{\boldmath $r$}\left\{
\phi^*\nabla\phi\right\}$ is ring momentum, and $E$ and 
$\mbox{\boldmath $P$}$ are measured in units of $\hbar n_0c\xi^2$ 
and $\hbar n_0\xi^2$, respectively. The energy as a function of flow velocity
for different obstacle heights is shown in Fig.~\ref{fig:5}.
With no object ($V=0$), Fig.~\ref{fig:5}(upper left), the energy of the 
ring decreases with increasing velocity reaching a minimum at $v=0.93$, 
which coressponds to the cusp in the dispersion curve (see inset). 
For $v>0.93$, the collapsed ring leaves a lower density, higher 
velocity region with energy $E\sim cP$, christened a 
rarefraction pulse by Jones and Roberts \cite{jon82}.  
Inserting $E= cP$ in Eq.~(\ref{eq:ener2}), one finds that
as $v$ decreases, $\vert\phi \vert^2$ must also decrease, but
this becomes impossible when $\vert\phi \vert^2=0$, so the rarefraction
pulse is replaced by a vortex ring. The process of supersonic flow
creating a localised sound wave which evolves into a vortex ring 
(or pair in 2D) appears to be central to the mechanism of vortex 
nucleation in dilute quantum fluids (see e.g. Fig.~4 in Ref.~\cite{jack98}).
For $V>0$, the energy of the laminar flow solution is no longer zero, and
the ring solution splits into two branches corresponding to 
the pinned ring and the encircling ring (Fig.~5 upper right). 
For low velocities (large radii), the energy of the encircling ring is
higher than the pinned ring by an amount corresponding to the energy 
of the ring segment inside the object. Note that the pinned and encircling 
ring solutions always merges below the critical velocity, Fig.~\ref{fig:5} 
lower right (inset). In a time-dependent simulation we observe 
a smooth transition (with no energy barrier) between laminar flow and 
the encircling ring supporting the quantum transition model of ring formation \cite{donn91,till90}. 

\section{Flow adjacent to a plane boundary}
\begin{figure}[htb]
\centering
\epsfig{file=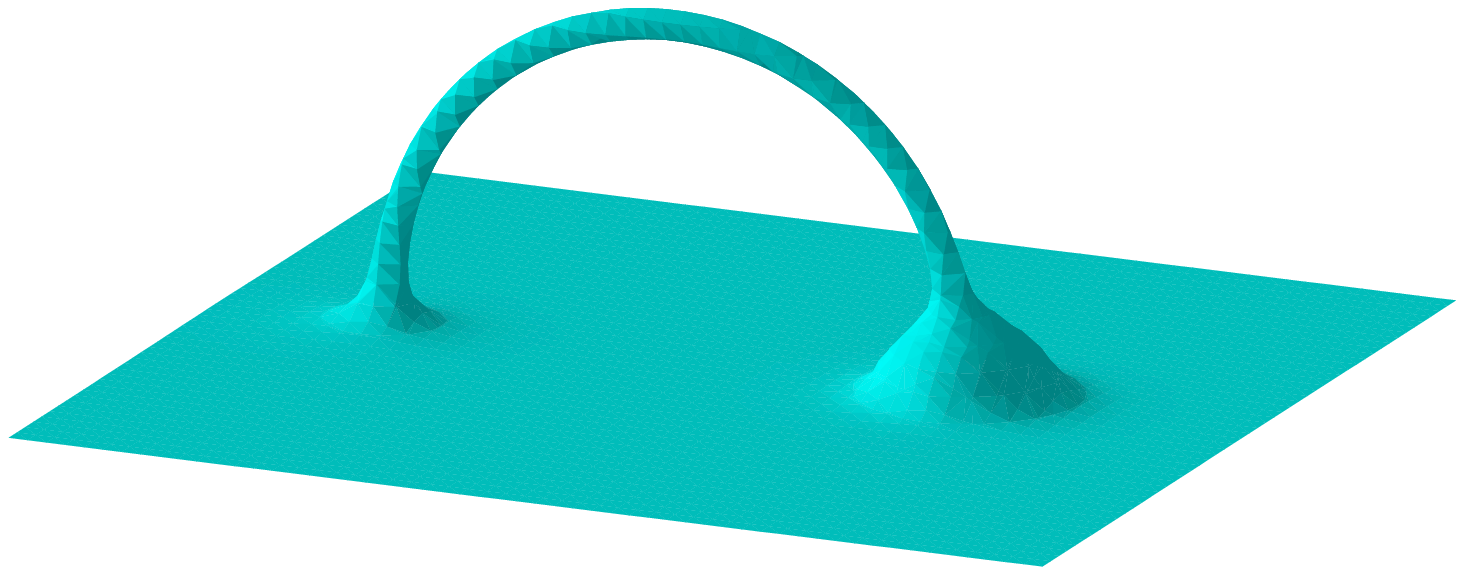,clip=,width=6cm,bbllx= 85,bblly=310,bburx=505,bbury=480}
\epsfig{file=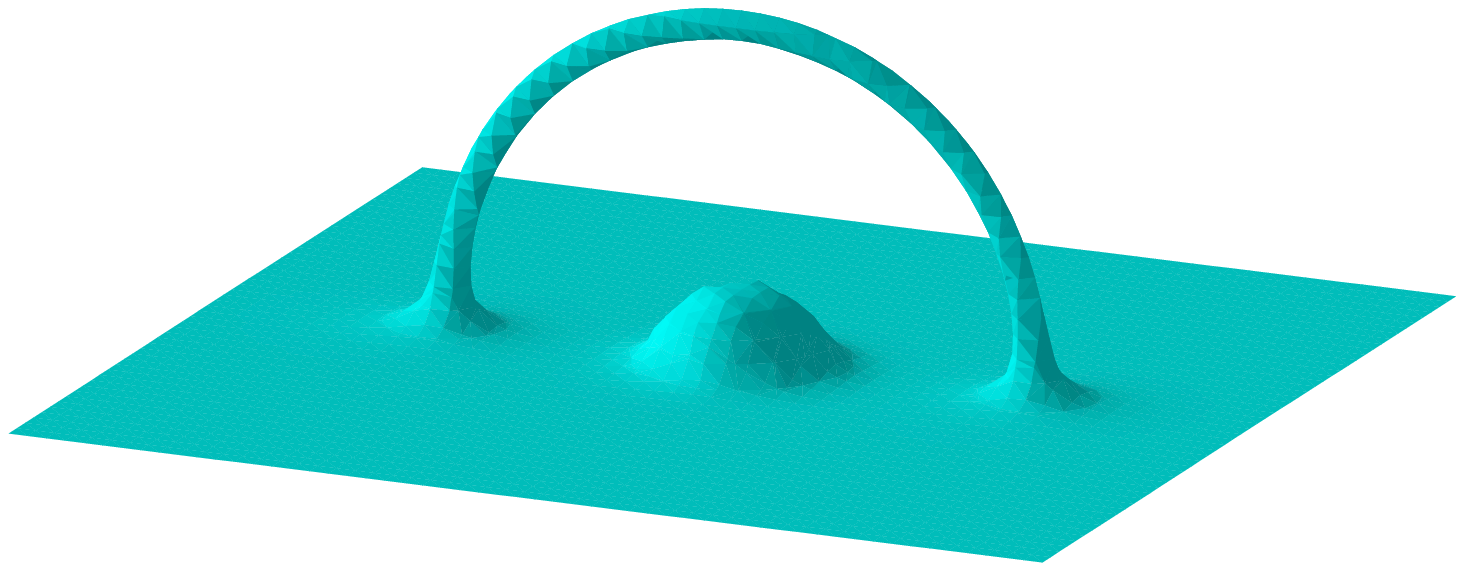,clip=,width=6cm,bbllx= 85,bblly=310,bburx=505,bbury=480}
\\
\epsfig{file=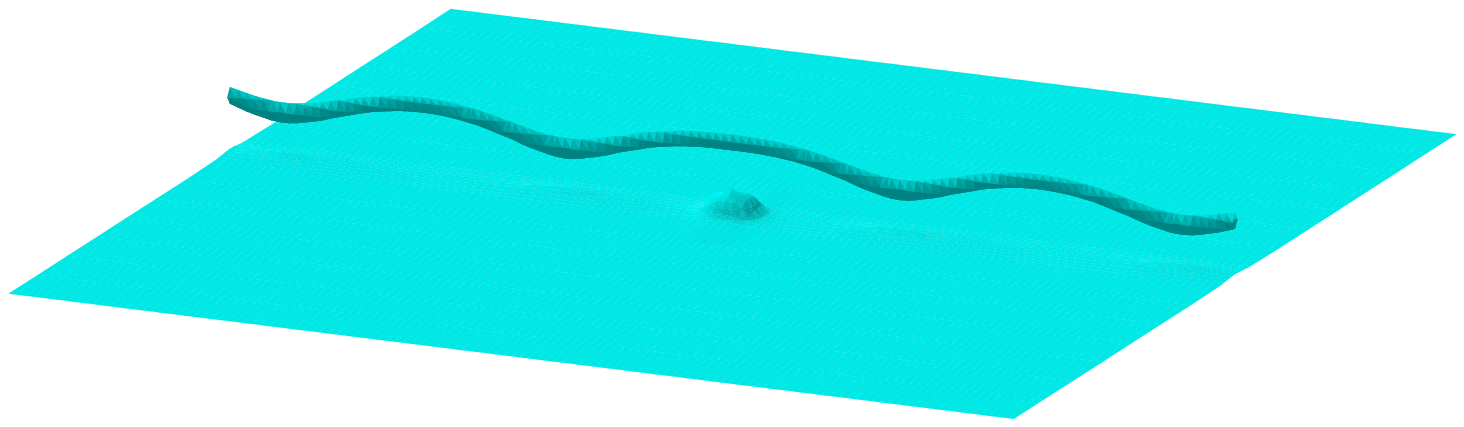,clip=,width=6cm,bbllx= 85,bblly=210,bburx=505,bbury=480}
\epsfig{file=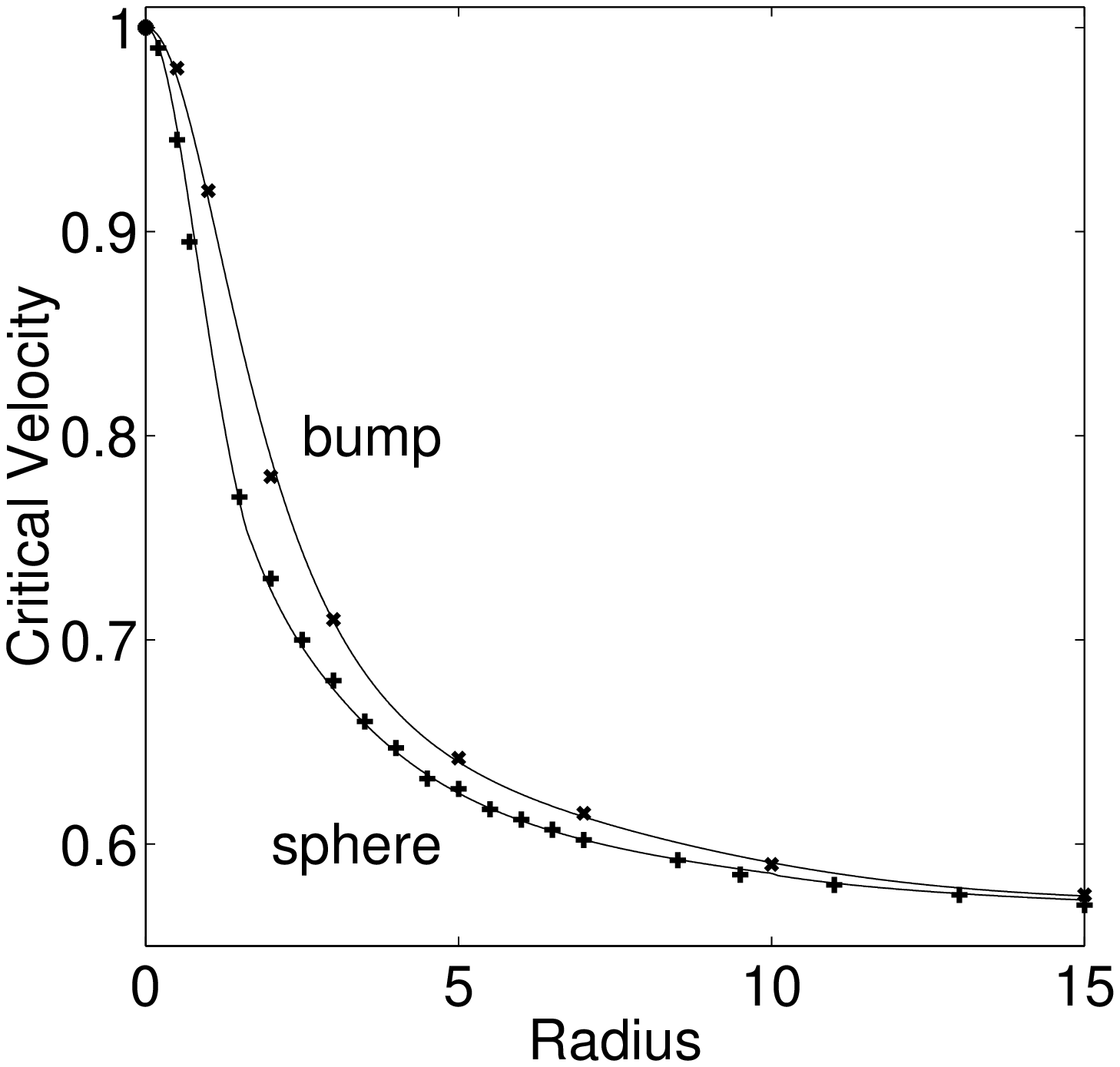,clip=,width=6cm,bbllx=50,bblly=186,bburx=535,bbury=595}
\caption{Surface density plots illustrating the stationary vortex structures
near a bump on a plane surface: vortex loops pinned or encircling the bump
(upper left and right) and a vortex line above the bump (lower left). 
The critical velocity as a function of the defect radius, $R$, for 
a hemishphere on a plane and a sphere in the bulk are shown
lower right. For large $R$, $v_{\rm c}\rightarrow 0.55$.}
\label{fig:6}
\end{figure}

For flow adjacent to a plane boundary, three classes of solution may be 
distinguished: laminar flow; vortex loops, and vortex lines parallel to 
the plane. The loop behaves similar to a free ring, i.e., the radius 
decreases with increasing velocity, and merges into the plane at a 
critical velocity, $v =1$. To comment on the flow of superfluids in 
real systems, we consider the effect of a surface bump. In this 
case, the vortex loop can either encircle or pin to the bump (again 
the pinned loop has a lower energy), and the vortex line acquires 
undulations, whose wavelength decreases with increasing flow velocity  
(Fig.~\ref{fig:6}).
The key effect of the bump is to reduce the 
critical velocity, $v_{\rm c}$ (Fig.~\ref{fig:6} lower right). In the limit 
of large radius, $v_{\rm c}\rightarrow 0.55$ for both surface and
volume defects. As $v_{\rm c}$ coincides with the appearance of drag in superfluids,
one may conjecture that surface roughness is a significant factor 
in determining the dissipation at low flow velocities. 

\section{Conclusions}
We have investigated vortex line and ring structures in dilute quantum fluids. 
We find that vortex tubes with multiple circulation consist of separate minima, 
and that near an object or surface lower energy, pinned solutions exist. 
Both laminar flow and vortex solutions become unstable at a critical velocity 
which is equal to the speed of sound for an unbounded flow or adjacent to a wall, 
but decreases near an object or surface bump. As the critical velocity corresponds 
to a transition between normal and drag-free flow, this dependence indicates 
how surface roughness can produce a marked effect on the flow of superfluids.

\stars Financial support for this project was provided by the EPSRC. 
TW is supported by the Studienstiftung des Deutschen Volkes.

\end{document}

%% file: euromacr.tex
\def\And{{\rm and\ }}

\def\stars{\bigskip\centerline{***}\medskip}

\newif\ifboo \boofalse


%% file: ring.bbl
\begin{thebibliography}{99}

\bibitem{donn91}
R. J. Donnelly, {\it Quantized vortices in Helium II}, (CUP, Cambridge,
1991).

\bibitem{till90}
D. R. Tilley and J. Tilley, {\it Superfluidity and superconductivity},
3rd Ed., (IoP, Bristol, 1990).


\bibitem{hind95}
M. B. Hindmarsh and T. W. B. Kibble, Rep. Prog. Phys. {\bf 58}, 477 (1995).

\bibitem{gpe} 
V. L. Ginzburg and L. P. Pitaevskii, Sov. Phys. JETP {\bf 7},
858 (1958); E. P. Gross, J. Math. Phys. {\bf 4}, 195 (1963). 

\bibitem{dalf99} F. Dalfovo, S. Giorgini, L. P. Pitaevskii, and S. Stringari,
Rev. Mod. Phys. {\bf 71}, 463 (1999).

\bibitem{hak97}
V. Hakim, Phys. Rev. E, {\bf 55}, 2835 (1997).

\bibitem{huep97}
C. Huepe and M.-\'E. Brachet, C. R. Acad. Sci. Paris,
{\bf 325}, 195 (1997).

\bibitem{wini99} 
T. Winiecki, J. F. McCann, and C. S. Adams, Phys. Rev. Lett. {\bf 82}, 5186 (1999).

\bibitem{numrec}
W. H. Press, S. A. Teukolsky, W. T. Vetterling and B. P. Flannary, 
{\it Numerical recipes in FORTRAN : the art of scientific computing}
2nd Ed. {CUP, Cambridge, 1992}.

\bibitem{aran96}
I. Aranson, and V. Steinberg, Phys. Rev. B {\bf 53}, 75 (1996).

\bibitem{rob71} P. H. Roberts and J. Grant, J. Phys. A {\bf 4}, 55 (1971).

\bibitem{jon82} C. A. Jones and P. H. Roberts, J. Phys. A {\bf 15}, 2599 (1982).

\bibitem{jack98} B. Jackson, J. F. McCann, and C. S. Adams,
Phys. Rev. Lett. {\bf 80}, 3903 (1998).

\end{thebibliography}
